\documentclass{article}
\usepackage[a4paper,left=1in,right=1in,top=3cm,bottom=2cm]{geometry}

\usepackage{graphicx}
\usepackage{amsmath}
\usepackage{algorithmic}
\usepackage{algorithm}

\usepackage{booktabs} 
\usepackage{array}

\usepackage[para,online,flushleft]{threeparttable}
\usepackage{diagbox}

 \usepackage{mathrsfs}

\begin{document}
\title{Generative Multi-Form Bayesian Optimization}

\author{Zhendong~Guo,
        ~Haitao~Liu,
        ~Yew-Soon~Ong,~\textit{Fellow,~IEEE,}
        ~Xinghua~Qu,\\
        ~Yuzhe~Zhang,
        and~Jianmin~Zheng
\thanks{Z. Guo is with the Data Science and Artificial Intelligence Research Center, School of Computer Science and Engineering, Nanyang Technological University, Singapore 639798. (e-mail: ericzhendong@163.com)}
\thanks{H. Liu is with the School of Energy and Power Engineering, Dalian University of Technology,116024, Dalian, China (e-mail: htliu@dlut.edu.cn).}
\thanks{Y.S. Ong is with the Agency for Science, Technology and Research (A*STAR), Singapore. (e-mail: oyewsoon@gmail.com)}
\thanks{X. Qu is with the Computational Intelligence Lab, School of Computer Science
and Engineering, Nanyang Technological University, Singapore, 639798
(email: xinghua001@e.ntu.edu.sg)}
\thanks{Y. Zhang and J. Zheng are with the School of Computer Science and Engineering, Nanyang Technological University, Singapore, 639798 (e-mail: \{zhang.yz,  ASJMZheng\}@ntu.edu.sg).}}

\markboth{}%
{Shell \MakeLowercase{\textit{et al.}}: Bare Demo of IEEEtran.cls for IEEE Journals}

\date{}
\maketitle

\begin{abstract}

Many real-world problems, such as airfoil design, involve optimizing a black-box expensive objective function over complex structured input space (e.g., discrete space or non-Euclidean space). By mapping the complex structured input space into a latent space of dozens of variables, a two-stage procedure labeled as generative model based optimization (GMO) in this paper, shows promise in solving such problems. However, the latent dimension of GMO is hard to determine, which may trigger the conflicting issue between desirable solution accuracy and convergence rate. To address the above issue, we propose a multi-form GMO approach, namely generative multi-form optimization (GMFoO), which conducts optimization over multiple latent spaces simultaneously to complement each other. More specifically, we devise a generative model which promotes positive correlation between latent spaces to facilitate effective knowledge transfer in GMFoO.
And further, by using Bayesian optimization (BO) as the optimizer, we propose two strategies to exchange information between these latent spaces continuously. 
Experimental results are presented on airfoil and corbel design problems and an area maximization problem as well to demonstrate that our proposed GMFoO converges to better designs on a limited computational budget.

\end{abstract}


\section{Introduction}
 
Many real-world problems, such as airfoil~\cite{li2020efficient} or hull-form shape design~\cite{2021Hull} involve optimizing a black-box expensive objective function over complex structured input space.
According to~\cite{tripp2020sample}, the structured input space refers to discrete space~\cite{chiu2020airfoil},~\cite{du2020b} or non-Euclidean space~\cite{kusner2017grammar},~\cite{2017Constrained}, which has the following characteristics, i.e., (1) the number of variables in the structured input space are 100+, and (2) the variables in the original structured input space can be highly interacted.
Hence, solving such kind of problems is challenging.

One promising way to tackle the above challenging problems is a two-stage procedure that has emerged over the past few years~\cite{chen2019aerodynamic},~\cite{lu2018structured},~\cite{nguyen2016synthesizing},~\cite{luo2018neural}, which is labeled as generative model based optimization (GMO) in this paper.
The general process of GMO is as follows. 
First, a generative model $g:\mathcal{Z} \mapsto \mathcal{X}$
that maps the structured input space $\mathcal{X}$ into a continuous latent space $\mathcal{Z}$ with dozens of variables, is built.
Then, optimization is carried out over this learnt latent space $\mathcal{Z}$.
Following this way, the original challenging problem becomes more accessible.

Though a series of successful stories has been reported~\cite{yang2018microstructural},~\cite{gomez2018automatic}, the effectiveness of GMO can be influenced by the latent dimension of generative model~\cite{chiu2020airfoil,du2020b}.
That is, when the latent dimension is set high, the convergence rate of GMO can drop dramatically with the increase of latent dimension, which is also known as the issue of ``curse of dimensionality"~\cite{shan2010survey}.
On the contrary, if the latent dimension is set low, the solution accuracy in such low-dimensional latent space can be limited, resulting in sub-optimal final solutions~\cite{chiu2020airfoil}.  
In other words, there is a balance between desirable solution accuracy and convergence rate when selecting the dimension of latent space for GMO.

To address the above balance issue, we propose a multi-form optimization approach, namely \emph{generative multi-form optimization} (GMFoO).
In particular, in contrast to the existing efforts that consider single latent space, we propose to include multiple latent spaces in GMFoO. 
Moreover, we propose to carry out optimizations over multiple latent spaces at the same time so as to benefit from each other.
The above idea is inspired by the emerging topic known as transfer optimization (TO)~\cite{gupta2017insights}, which attempts to gain knowledge from the related tasks~\cite{da2018curbing},~\cite{zhou2020toward},~\cite{liang2020evolutionary},~\cite{li2021meta} or alternate formulations~\cite{2016Evolutionary},~\cite{zhang2021study} to achieve better solutions with even fewer computational costs.
In particular for our proposed GMFoO, the quicker optimization progress of the low-dimensional latent space can help to accelerate the process of the high-dimensional latent space.
In turn, the better solutions in the high-dimensional latent space may help to resolve the solution accuracy issue in low-dimensional latent space.
Thereby, GMFoO can take a balance in between desirable solution accuracy and convergence rate to achieve the optimal solutions efficiently.

Also note that, various algorithms such as evolutionary algorithms (EA)~\cite{peng2018multimodal},\cite{cai2019efficient}, particle swarm optimization (PSO) algorithms~\cite{cheng2015social},~\cite{wang2017committee} can be used in GMO for the optimization over latent space.
Among them, Bayesian optimization (BO)~\cite{2018A},~\cite{Martinez2019Funneled} gains particular attentions in the GMO paradigm~\cite{2020Good}, since it is well-known to be sample-efficient for solving expensive black-box problems~\cite{2021Guo},~\cite{2020Generalizing}.
Therefore, we study GMFoO with BO in this paper.

The main contribution of this paper can be summarized as follows:



\begin{itemize}

    \item We propose the novel GMFoO framework, for solving black-box problems that involve complex structured input space. Instead of optimizing in single latent space as most GMO studies do, the GMFoO conducts optimization over multiple latent spaces simultaneously to benefit from each other, therefore achieving a balance between desirable solution accuracy and convergence rate.
    
   

    
    \item Particularly, we devise a generative model that promotes positive correlations among latent spaces in the training stage to facilitate effective knowledge transfer in GMFoO. And further, we instantiates the proposed GMFoO with BO, where two strategies are proposed to exchange information between the latent spaces continuously.
    \item Through tests on the airfoil and corbel design problems and an area maximization problem, the effectiveness of GMFoO has been well demonstrated.
\end{itemize}

The remainder of this paper is organized as follows. Section II presents the preliminaries.
And then, Section III illustrates the details of GMFoO framework and the related algorithm. After that, Section IV shows the experimental studies.
Finally, we draw conclusions in Section V.

\section{Preliminaries}


In order to outline the contributions of this work more clearly, 
we provide preliminaries on some related topics in this section, such as generative model based optimization, multi-form optimization and Bayesian optimization.  


\subsection{Generative Model-based Optimization}

\begin{figure*}
	\centering
	\includegraphics[width=5.5in]{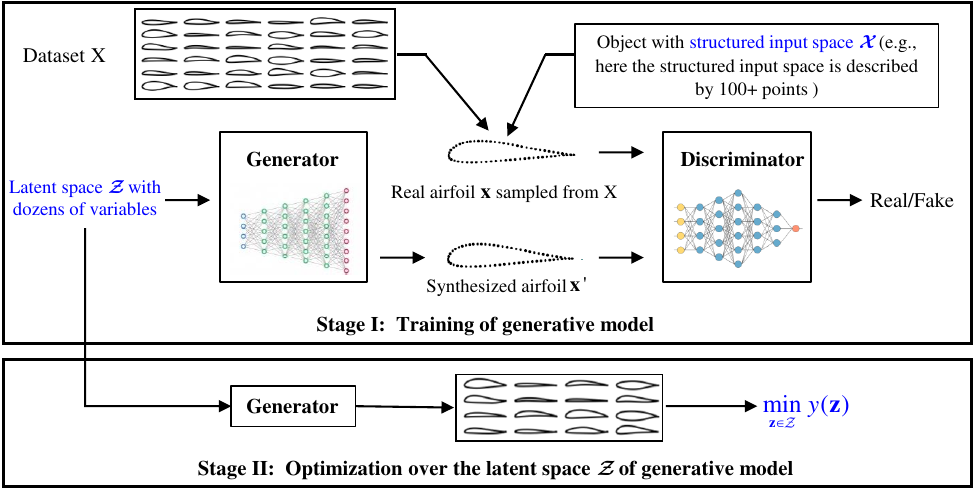}
	\caption{Standard process of generative model based optimization (GMO)}
\end{figure*}

The GMO paradigm can be formulated as:
\begin{equation}
\begin{aligned}
\mathop {\min }\limits_{{\bf{x}} \in {\rm{{\cal X}}}} f({\bf{x}}) &= \mathop {\min }\limits_{{\bf{z}} \in {\rm{{\cal Z}}}} f(g({\bf{z}}))\\
 &\approx \mathop {\min }\limits_{{\bf{z}} \in {\rm{{\cal Z}}}} y({\bf{z}})
\end{aligned}
\end{equation}
where, $\mathcal{X}$ is the structured input space, $\mathcal{Z}$ is the latent space learnt by using a generative model $g$, and $y$ is the latent objective model to approximate the objective function $f$.

While various generative models such as variational autoencoder (VAE)~\cite{kingma2013auto} can be employed, here the generative adversarial net (GAN)~\cite{goodfellow2014generative},~\cite{creswell2018generative} is used to illustrate the process of mapping the structured input space into latent space. 
As shown in Fig.1, GAN is consisted of two components as the generator and the discriminator. 
With the input of latent vector $\bf{z}$, the generator is trained to synthesize the structured object, the discriminator is trained to distinguish the synthesized objects $\bf{x'}$ (also denoted by $G(\bf{z})$ in Eq.(2)) from the real data $\bf{x}$, with the following adversarial training loss:
\begin{equation}
\begin{aligned}
 \min_G \max_D V\left( D,G \right) &= E_{{\bf{x}}\sim P_{data}}\left[ \log D\left(\bf{x} \right) \right]\\
 &+ E_{{{\bf{z}}}\sim P_{\bf{z}}}\left[\log \left( 1 - D\left( G\left( \bf{z} \right) \right) \right) \right]
\end{aligned}
\end{equation}
where, $D(\bf{x})$ and $D(G(\bf{z}))$ are the probabilities that the samples coming from the real dataset $X$ or the generator, respectively. 
After training, the generator of GAN can synthesize various $\bf{x}$ by varying ${\bf{z}}$. Then, optimization can be carried out over this learnt latent space with latent vector $\bf{z}$.
Correspondingly, the related pseudo code of standard GMO procedure is summarized in Algorithm 1.

Thus far, many successful applications of GMO has been reported~\cite{chiu2020airfoil},~\cite{2017Constrained},~\cite{chen2019aerodynamic},~\cite{lu2018structured},~\cite{nguyen2016synthesizing},~\cite{luo2018neural}.
However, further work is required to improve the performance of GMO. 
For instance, as discussed in the Introduction part, the selection of latent dimension can greatly influence the GMO performance, which 
may trigger the conflicting issue between desirable solution accuracy and convergence rate.
To this end, the key contribution of this work is that, \emph{instead of optimizing in single latent space as most GMO approaches do, we propose a multi-form approach, which conducts optimization in multiple latent spaces at the same time to address the above issue}.

\begin{algorithm}[t]
	\caption{Generative Model based Optimization (GMO)}
	\begin{algorithmic}[1]
		\REQUIRE{Target optimization problem, structured very high-dimensional dataset \begin{small}$X$\end{small}}
		\ENSURE{Optimal solution of the target problem}
		\STATE {\bf{Training of the Latent Space:}} Train a latent representation \begin{small}${\bf{z}} \in R^d$\end{small} for data \begin{small}${\bf{x}} \in R^D$\end{small} (\begin{small}$d \ll D$\end{small}) by using generative models (e.g., GAN) with the dataset \begin{small}$X$\end{small}, and thereby embedding the structured high-dimensional input space \begin{small}$\mathcal{X}$\end{small} into the much lower-dimensional latent space \begin{small}$\mathcal{Z}$\end{small}

		\STATE {\bf{Optimization in the Latent Space:}} Use algorithms such as Bayesian optimization, genetic algorithms, etc. to carry out optimization in the latent space.
	\end{algorithmic}
\end{algorithm}

\subsection{Transfer Optimization and Multi-Form Optimization}


Instead of starting the optimization search from scratch as most traditional approaches do, transfer optimization (TO) proposes to gain knowledge from the related tasks to achieve better solutions with even fewer computational costs~\cite{min2017knowledge},~\cite{yogatama2014efficient}. 
Recently, several comprehensive and insightful reviews of TO have been reported~\cite{gupta2017insights},~\cite{tan2021evolutionary},~\cite{wei2021review}.
According to~\cite{gupta2017insights}, TO can be classified into three categories, i.e., sequential transfer optimization (STO)~\cite{da2018curbing},~\cite{joy2019flexible}, multitasking optimization (MTO)~\cite{lin2019multi},~\cite{swersky2013multi} and multi-form optimization (MFoO)~\cite{zhang2021study}.
While STO and MTO deal with distinct optimization tasks that are presumed to be related, MFoO exploits alternate formulations of a single target task.

More formally, in hope to improve the efficiency of an search algorithm, MFoO proposes joint optimization of several alternate formulations, with continuous knowledge transfer between them:
\begin{equation}
\begin{small}
T = \mathop {\min }\limits_{{\bf{x}} \in {\rm{{\cal X}}}} f({\bf{x}})
\end{small}
\end{equation}
\begin{equation}
\begin{small}
Q_t^{MFoO}(T|{T_1}, \cdots ,{T_K},M(t)) - Q_t(T) \ge 0
\end{small}
\end{equation}
where, $T$ denotes the target optimization task, $Q_t$ is a measure quantifying the quality of solution(s) obtained at the $t^{th}$ iteration, ${T_1}, \cdots ,{T_K}$ are alternate formulations of $T$, and $M(t)$ is the knowledge base to facilitate knowledge transfer between the alternate formulations.

The previous efforts of MFoO in building alternate formation $T_k$ can be mainly categorized into two groups.
The first group of studies propose to build $T_k$ by reformulating the objective function as $f_k$:
\begin{equation}
\begin{small}
{T_k} = \mathop {\min }\limits_{{\bf{x}} \in {\rm{{\cal X}}}} {f_k}({\bf{x}},M(t))
\end{small}
\end{equation}
For instance, some works attempt to multiobjectivize a single-objective problem of interest to remove the local optima~\cite{2016Evolutionary},~\cite{2001Reducing},~\cite{2008Multiobjectivization}, some others propose to augment the expensive high-fidelity objective evaluations by a large number of low-fidelity but cheap samples to accelerate the optimization progress~\cite{2021Parallel},~\cite{guo2018analysis},~\cite{li2020multi}.

The second group of MFoO studies propose to build $T_k$ by constructing alternate searching space ${{\rm{{\cal X}}}_k}$:
\begin{equation}
\begin{small}
{T_k} = \mathop {\min }\limits_{{\bf{x}} \in {{\rm{{\cal X}}}_k}} f_k({{\bf{x}}_k},M(t))
\end{small}
\end{equation}
For example, some work proposes to carry out multi-form optimization with both constrained and unconstrained formulations~\cite{2002Theoretical}.
Alternatively, some studies proposes to divide the full design space into subset spaces. Then, optimizations are conducted over subset spaces by fixing the remaining variables as constant as those of the current best solution~\cite{2018Nash}.


The work in this paper belongs to the second group of MFoO shown in Eq.(6).
Compared to the existing studies, the distinction of this work is that, \emph{we use a generative model to promote positive correlation between the alternate searching spaces in our work.
And accordingly, the optimizations in these correlated searching spaces can complement each other to achieve the optimal solution efficiently.}

\subsection{Bayesian Optimization (BO)}

Due to its sample efficiency~\cite{shahriari2015taking}, BO is most frequently used as the optimizer in GMO~\cite{chiu2020airfoil},~\cite{yang2018microstructural},~\cite{gomez2018automatic},~\cite{2020Good}. 
The general procedure of BO is shown in Algorithm 2. 
First, it uses Gaussian process (GP)~\cite{rasmussen2010gaussian} to build a surrogate of the objective function.
Second, it employs an acquisition function
which incorporates the GP surrogate to select the next promising solution candidates to sample, and simulations are conducted to evaluate the new sample candidate.
Third, the new sample is added to the training set for the next iteration.

As a key component of BO, GP is often used to approximate the input-output relation, which formulates the function prediction $Y({\bf{x}})$ as a Gaussian random variable:
\begin{equation}
\begin{scriptsize}
Y({\bf{x}}) \sim {\rm{GP}}(m({\bf{x}}),k({\bf{x}},{\bf{x}}'))
\end{scriptsize}
\end{equation}
where $m({\bf{x}})$ denotes the mean function, and $k({\bf{x}},{\bf{x}}')$ is the covariance function between samples $\bf{x}$ and ${\bf{x}}'$.
For standard GP with samples from a single source, 
the posterior estimates of the mean function prediction ($\hat y$)  and the corresponding mean squared error (${\hat \sigma ^2}$) at a point ${{\bf{x}}}$ are expressed as:
\begin{equation}
\begin{small}
\begin{aligned}
\hat y({{\bf{x}}}) &= {\bf{k}}({{\bf{x}}},X){\left( {K + \sigma _n^2{\bf{I}}} \right)^{ - 1}}{\bf{y}}\\
{{\hat \sigma }^2}({{\bf{x}}}) &= {\bf{k}}({{\bf{x}}},{{\bf{x}}}) - {\bf{k}}({{\bf{x}}},X){\left( {K + \sigma _n^2{\bf{I}}} \right)^{ - 1}}{\bf{k}}(X,{{\bf{x}}}) + \sigma _n^2
\end{aligned}
\end{small}
\end{equation}
where, $\sigma _n^2$ is the Gaussian noise, $K$ is the covariance matrix over the input features of samples $X$, and ${\bf{k}}$ is the cross-correlation vector between ${{\bf{x}}}$ and $X$.

\begin{algorithm}[t]
	\caption{Standard Bayesian Optimization (BO)}
	\begin{algorithmic}[1]
		\REQUIRE{Target optimization problem}
		\ENSURE{Optimal solution of the target problem}
		\STATE {\bf{Design of Experiment (DoE):}} Use space-filling technique such as Latin hypercube sampling to generate the initial samples and evaluate to obtain the pairwise labeled training set \begin{small}$\{X, Y\}$\end{small}
		\WHILE{the termination condition is not satisfied}
		\STATE {\bf{Build or Update GP Surrogate:}} Build GP surrogate with training set \begin{small}$\{X, Y\}$\end{small}
		\STATE {\bf{Sample Search with Acquisition function:}} Use acquisition function such as EI to find the next sample to query, e.g., \begin{small}${{\bf{x}}^*} = \arg \mathop {\max }\limits_{\bf{x}} EI({\bf{x}})$\end{small}
		\STATE {\bf{Update the Training Set:}} Evaluate the label $y^*$ of the new sample \begin{small}$\bf{x^*}$\end{small} and append \begin{small}$\{X, Y\}$\end{small} with \begin{small}$\{{\bf{x^*}}, y^*\}$\end{small}
		\ENDWHILE
	
	\end{algorithmic}
\end{algorithm}

As another key component of BO, the acquisition function acts as a surrogate that determines which sample to be evaluated next.
Among various acquisition functions such as probability of improvement (PI)~\cite{shahriari2015taking}, upper-bound confidence (UCB)~\cite{2013Parallel}, etc., expected improvement (EI)~\cite{jones1998efficient} is most frequently used. 
Assuming that the function prediction  $Y({\bf{x}})$ comes from a GP with the posterior estimate as $Y({\bf{x}}) \sim N(\hat y({\bf{x}}),{\sigma ^2}({\bf{x}}))$, the basic idea of EI is to measure how much objective improvement can be attained with respect to the current best solution ${f_{\min }}$, which is formulated as:
\begin{equation}
\begin{small}
\begin{aligned}
EI({\bf{x}}) &=  {(f_{\min }^{} - \hat y({\bf{x}}))\Phi \left( u \right)} +  {\sigma ({\bf{x}})\phi \left( u \right)}\\
u &= {{({f_{\min }} - \hat y({\bf{x}}))} \mathord{\left/
		{\vphantom {{({f_{\min }} - \hat y({\bf{x}}))} {\sigma ({\bf{x}})}}} \right.
		\kern-\nulldelimiterspace} {\sigma ({\bf{x}})}}
\end{aligned}
\end{small}
\end{equation}
where, $\Phi \left(  \cdot  \right)$ and $\phi \left(  \cdot  \right)$ denote the standard normal distribution function and density function.
In each optimization cycle, we select the next point to sample by maximizing the EI:
\begin{equation}
\begin{small}
{{\bf{x}}^*} = \arg \mathop {\max }\limits_{\bf{x}} EI({\bf{x}})
\end{small}
\end{equation}

Note that most BO algorithms work well for optimization over domains of moderate dimensions~\cite{2018A},~\cite{Martinez2019Funneled}. 
And accordingly, the effectiveness of GMO can become questionable when the dimension of latent space is set too large (e.g., larger than 15).
From this end, \emph{the highlight of this work is that, we propose a multi-form approach that makes use of multiple searching spaces to help resolve the limitations of BO.}



With the above, it is not hard to comprehend that this work aims to solve the black-box problems that involve complex structured input space, by proposing a novel GMO approach.
In particular, instead of optimizing in single latent space as most GMO studies do, we propose a multi-form GMO approach, which conducts optimization in multiple latent spaces at the same time to achieve a balance between convergence rate and desirable solution accuracy. 

\section{Methodology}

In this section, we first show the general framework of GMFoO, then we instantiates GMFoO with BO.

\subsection{GMFoO framework}
\begin{figure*}
	\centering
	\includegraphics[width=5.5in]{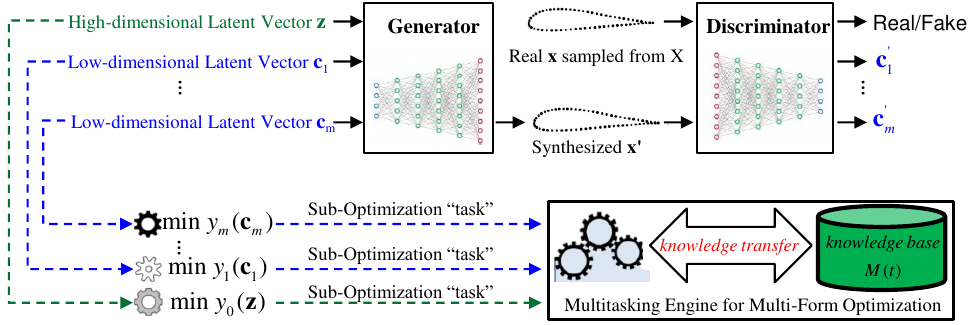}
	\caption{GMFoO framework, where the generative model is labeled as MFoO-GAN}
\end{figure*}

\begin{figure*}
	\centering
	\includegraphics[width=5.5in]{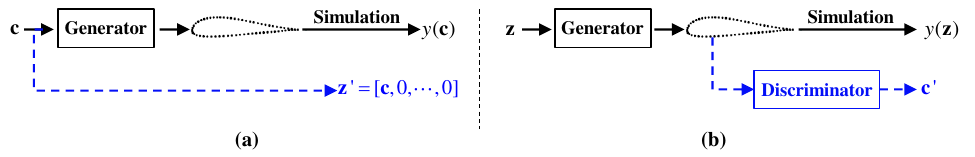}
	\caption{Sample synthesis and exchange process between the high- and lower-dimensional latent space, (a) in the low-dimensional latent space, (b) in the high-dimensional latent space}
\end{figure*}
\begin{figure*}[t!]
	\centering
	\includegraphics[width=5.5in]{"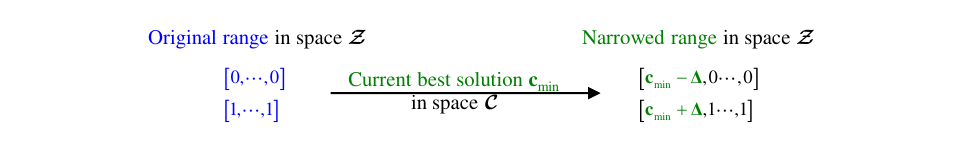"}
	\caption{Enhanced local exploitation of the high-dimensional latent space by leveraging information of the optimal solution of the low-dimensional latent space}
\end{figure*}
Given that a set of latent spaces are learnt by using a generative model, the GMFoO framework can be formulated as:
\begin{equation}
\begin{aligned}
\mathop {\min }\limits_{{\bf{x}} \in {\rm{{\cal X}}}} f({\bf{x}}) = \min \{ {T_0},{T_1}, \cdots ,{T_m}|M(t)\} 
\end{aligned}
\end{equation}
where, $\rm{{\cal X}}$ is the structured input space of a black-box problem, ${T_0},{T_1}, \cdots, {T_m}$ are the ``sub-tasks" that conduct optimization in one of the learnt latent space, and $M(t)$ is the knowledge base to facilitate information exchange among the ``sub-tasks".
Furthermore, in order to achieve a balance between desirable solution accuracy and convergence rate, we propose to include one high-dimensional latent space ${\rm{{\cal Z}}}$ and one or several low-dimensional latent spaces ${{\rm{{\cal C}}}_1}, \cdots ,{{\rm{{\cal C}}}_m}$ in the generative model $g$. And accordingly, the ``sub-tasks" in GMFoO can be expressed as:
\begin{equation}
\begin{aligned}
{T_0} &= \min f(g({\bf{z}})) \approx \mathop {\min }\limits_{{\bf{z}} \in {\rm{{\cal Z}}}} y_0({\bf{z}})\\
{T_i} &= \min f(g({{\bf{c}}_i})) \approx \mathop {\min }\limits_{{{\bf{c}}_i} \in {{\rm{{\cal C}}}_i}} y_i({{\bf{c}}_i}),\;\;\;
i= 1, \cdots, m
\end{aligned}
\end{equation}

Three questions may arise when instantiating the GMFoO framework. That is, (1) how to train ${\rm{{\cal Z}}}$ and ${{\rm{{\cal C}}}_1}, \cdots ,{{\rm{{\cal C}}}_m}$ to be correlated so as to promote effective knowledge transfer; 
(2) how to exchange samples between ${\rm{{\cal Z}}}$ and ${{\rm{{\cal C}}}_1}, \cdots ,{{\rm{{\cal C}}}_m}$; 
and (3) how to build $M(t)$ to help the ``sub-tasks" to benefit from each other.
The questions (1) and (2) mainly concerns how to build the generative model, which will be discussed in this subsection. And (3) is related to the proposed optimization algorithm, which will be discussed in the next subsection.



Figure 2 shows the GMFoO framework we proposed to answer the questions (1) and (2), where 
the generative model is labeled as MFoO-GAN, i.e., GAN for generative multi-form optimization.
More specifically, to promote positive correlations among $\mathcal{Z}$ and ${\mathcal{C}}_i$, we follow the idea of Info-GAN~\cite{chen2016infogan} by imposing a regularization term of maximizing the mutual information $I\left( {{\bf{c}}_i,G({\bf{z}})} \right)$ in the training loss:
\begin{equation}
\begin{small}
\min_G \max_D V_M\left( D,G \right) = V(D,G) - \sum\limits_{i = 1}^m {{\lambda _i}I\left( {{\bf{c}}_{i},G\left( \bf{z} \right)} \right)}
\end{small}
\end{equation}
where, ${\lambda _i}$ is the weight of a related regularization term.
For simplicity, we use $\bf{c}$ and $\bf{c}^{'}$  to denote ${\bf{c}}_i$ and ${\bf{c}}_i^{'}$, respectively, in the following paragraphs. Let $P({\bf{c}}|\bf{x})$ denote the real probabilistic distribution of the inverse inference of $\bf{x}$, and $Q({\bf{c}}|\bf{x})$ a factored Gaussian distribution that used to approximate $P({\bf{c}}|\bf{x})$, the mutual information $I\left( {{\bf{c}},G({\bf{z}})} \right)$ can be derived as:
\begin{equation}
\begin{small}
\begin{aligned}
I({\bf{c}};G({\bf{z}})) &= H({\bf{c}}) - H({\bf{c}}|G({\bf{z}}))\\
 &= {E_{{\bf{x}} \sim G({\bf{z}})}}\left[ {{E_{\bf{\tilde c} \sim P({\bf{c}}|{\bf{x}})}}\left[ {\log P({\bf{c}}|{\bf{x}})} \right]} \right] + H({\bf{c}})\\
 &= E_{{\bf{x}} \sim G({\bf{z}})}^{}\underbrace {\left[ {{D_{KL}}\left( {P( \cdot |{\bf{x}})||Q( \cdot |{\bf{x}})} \right)} \right]}_{ \ge 0}\\
 &+ {E_{\tilde c \sim P({\bf{c}}|{\bf{x}})}}\left[ {\log Q({\bf{c}}|{\bf{x}})} \right] + H({\bf{c}})\\
 &\ge {E_{{\bf{x}} \sim G({\bf{z}})}}\left[ {{E_{\tilde c \sim P({\bf{c}}|{\bf{x}})}}\left[ {\log Q({\bf{c}}|{\bf{x}})} \right]} \right] + H({\bf{c}})\\
 &=  - \sum\limits_{j = 1}^{|{\bf{c}}|} {\left( {\frac{{{{\left( {{c_j} - {\mu _j}} \right)}^2}}}{{2\sigma _j^2}} + \log \left( {{\sigma _j}\sqrt {2\pi } } \right)} \right)}  + H({\bf{c}})\\
 &\equiv {L_I}\left( {{\bf{c}},G({\bf{z}})} \right)
\end{aligned}
\end{small}
\end{equation}
where, $H( \cdot )$ is the information entropy, and $\mu_j$ and $\sigma_j^2$ are the mean and standard derivation of a Gaussian distribution for the $j_{th}$ component of the latent variable $\bf{c}$.
Additionally, ${L_I}\left( {{\bf{c}},G({\bf{z}})} \right)$ is the lower bound of $I\left( {{\bf{c}},G({\bf{z}})} \right)$.

Since $I\left( {{\bf{c}}_i,G({\bf{z}})} \right)$ is only active when training the generator, we can 
pick out the regularization term from Eq.(13), with denoting $\mu_j = c_j^{'}$ and fixing $H(\bf{c})$ as a constant, as below:
\begin{equation}
\begin{small}
\begin{aligned}
\min -{L_I}\left( {{\bf{c}},G({\bf{z}})} \right) \simeq \;\;\min \sum\limits_{j = 1}^{|{\bf{c}}|} {\left( {\frac{{{{\left( {{c_j} - {c_j}'} \right)}^2}}}{{2\sigma _j^2}} + \log \left( {{\sigma _j}\sqrt {2\pi } } \right)} \right)}
\end{aligned}
\end{small}
\end{equation}
Then, by minimizing the gap between $\bf{c}$ and the inverse inference $\bf{c^{'}}$ through Eq.(15), the major feature of the the object $G(\bf{z})$ is captured by $\bf{c}$, and the low-dimensional latent space ${\mathcal{C}}$ is trained to be correlated with the high-dimensional latent space ${\mathcal{Z}}$.

For the sample exchange from the low-dimensional latent space ${\mathcal{C}}$ to the high-dimensional latent space ${\mathcal{Z}}$, we propose to fix the remaining variables (denoted by ${{\bf{z}}^*}$) of ${\bf{z}} = [{\bf{c}},{{\bf{z}}^*}]$ as constants, e.g., fixing ${{\bf{z}}^*}$ as zeros as shown in Fig.3(a).
The benefits of doing so can be as follows.
First, fixing ${{\bf{z}}^*}$ helps to prioritize the optimization search in directions of major variability of the object $\bf{x}$, and thus helping to accelerate the overall optimization progress.
Second, by fixing ${{\bf{z}}^*}$ instead of assigning it as a random noise, the samples in the latent space ${\mathcal{C}}$ can be reused in the next iteration, without worrying the negative influence caused by the randomness of ${{\bf{z}}^*}$. 
Additionally, fixing ${{\bf{z}}^*}$ makes it easy to transform samples from the low-dimensional latent space to the high-dimensional latent space.

Conversely, for the sample exchange from the high-dimensional latent space ${\mathcal{Z}}$ to the low-dimensional latent space ${\mathcal{C}}$, we propose to use the inverse inference structure of the discriminator, as shown in Fig.3(b). 
More specifically, ${\bf{c}}^{'}$ is the mean of the posterior inverse inference of $\bf{x}$ with the learnt probability $Q({\bf{c}}^{'}|\bf{x})$.
Compared to $\bf{c}$ as a prior component of ${\bf{z}} = [{\bf{c}},{{\bf{z}}^*}]$ to generate the object $\bf{x}$, ${\bf{c}}^{'}$ also takes the variability of $\bf{x}$ caused by ${{\bf{z}}^*}$ into account via $\sigma_j^2$ (see Eq.(14)) in the training process.
In other words, ${\bf{c}}^{'}$ can capture the major features of $\bf{x}$ with better robustness when doing the inverse inference.
Therefore, we propose to use ${\bf{c}}^{'}$ with the process shown in Fig.3(b) to transform samples from the high-dimensional latent space to the low-dimensional latent space.

\subsection{An instantiation of GMFoO with BO}

While different versions of GMFoO can be proposed by building different knowledge base $M(t)$, one possible instantiation is presented in this subsection.
With the focus on discussing how to build knowledge base $M(t)$ to help the ``sub-tasks" to benefit from each other, only one high-dimensional and one low-dimensional latent spaces are considered.

~\\
\indent
1. Enhanced Local Exploitation

As a subset of the high-dimensional latent space ${\mathcal{Z}}$, the low-dimensional latent space ${\mathcal{C}}$ of MFoO-GAN can capture the major variability of the target object.
It means, the current best solution of ${\mathcal{C}}$, i.e., ${\bf{c}}_{min}$, can be located in the neighborhood of the real optimal solution of ${\mathcal{Z}}$. In other words, ${\bf{c}}_{min}$ can be used as an indicator to enhance the local exploitation of the promising areas of ${\mathcal{Z}}$.
Taking the cue, we build knowledge base $M(t)$ as shown in Fig.4, where we use ${\bf{c}}_{min}$ to narrow the searching space of ${\mathcal{Z}}$. 
Considering that the neighborhood of the optimal solution of ${\mathcal{Z}}$ and ${\mathcal{C}}$ may not be exactly overlapped, an additional variable $\Delta $ is used to control the size of the narrowed space.
The influence of changing $\Delta $ will be discussed in Section IV.

~\\
\indent
2. GP-based Multi-Fidelity Optimization

Note that the sample efficiency of BO is highly related to the accuracy of the GP surrogate.
One direct way to improve the GP accuracy is to increase the training samples.
However, as discussed with Fig.3, there is accuracy loss when transforming the samples from the high-dimensional latent space ${\mathcal{Z}}$ to the low-dimensional latent space ${\mathcal{C}}$.
In the meantime, directly injecting the samples of ${\mathcal{C}}$ into ${\mathcal{Z}}$ may further enhance the exploitation of these major directions, but ignoring exploration along the remaining directions (i.e., ${\bf{z}}^*$ from ${\bf{z}}=[{\bf{c}}, {\bf{z}}^*]$).
In view of the above, we propose to treat the samples transformed from both ${\mathcal{Z}}$ and ${\mathcal{C}}$ as low-fidelity samples. 
And accordingly, we propose to build multi-fidelity GP (MFGP), which utilizes the covariance matrix as knowledge base $M(t)$ to extract information from the low-fidelity samples to improve the surrogate accuracy.

More specifically, we propose to build two MFGP surrogates to carry out optimization in GMFoO.
Let $C$ and $Z$ denote the high-fidelity samples of ${\mathcal{C}}$ and ${\mathcal{Z}}$, respectively, and $C^{'}$ and $Z^{'}$ the low-fidelity samples transformed from ${\mathcal{Z}}$ and ${\mathcal{C}}$, respectively.
In the meantime, ${\bf{y}}(C)$ and ${\bf{y}}(Z)$ are the vectors of objective function values of related samples.
Then, the correlation vector and covariance matrix of MFGP built in the low-dimensional latent space are formulated as:
\begin{equation}
\begin{small}
\begin{aligned}			
{{\bf{k}}_M}({\bf{c}},C^*)&=
\left[ {{\rho _{11}}{\bf{k}}({\bf{c}},{C}),{\rho _{12}}{\bf{k}}({\bf{c}},{C'})} \right]\\
{K_M}\left( {C^*} \right) &= \left[ {\begin{array}{*{20}{c}}
{{\rho _{11}}K\left( {C} \right)}&{{\rho _{12}}K\left( {C,C'} \right)}\\
{{\rho _{21}}K\left( {C',C} \right)}&{{\rho _{22}}K\left( {C'} \right)}
\end{array}} \right] + D
\end{aligned}
\end{small}
\end{equation}
where, $\bf{k}$ and $K$ are the correlation vector and matrix formulation of input variables (see Eq.(8));
${\rho _{ij}}$ denotes the correlation coefficient to model the relatedness of \begin{small}${\bf{y}}(C)$\end{small} and \begin{small}${\bf{y}}(Z)$\end{small};
$D$ is an $2 \times 2$ diagonal matrix with diagonal elements \begin{small}${\left\{ {\sigma _{n,j}^2} \right\}_{1 \le j \le 2}}$\end{small}, where $\sigma _{n,j}^2$ is the noise term associated with \begin{small}${\bf{y}}(C)$\end{small} or \begin{small}${\bf{y}}(Z)$\end{small}. 
Then, let \begin{small}$C^*=[C,C']$\end{small}, the prediction at an unobserved point ${{\bf{c}}}$ can be expressed as: 
\begin{equation}
\begin{aligned}	
\hat y\left( {\bf{c}} \right) &= {{\bf{k}}_M}\left( {{\bf{c}},{C^*}} \right)K_M^{ - 1}\left( {{C^*}} \right)\left[ \begin{array}{l}
{\bf{y}}\left( C \right)\\
{\bf{y}}\left( Z \right)
\end{array} \right]\\
{\sigma ^2}\left( {\bf{c}} \right) &= {{\bf{k}}_M}\left( {\bf{c}} \right) - {{\bf{k}}_M}\left( {{\bf{c}},{C^*}} \right)K_M^{ - 1}{{\bf{k}}_M}\left( {{C^*},{\bf{c}}} \right) + \sigma _{n,1}^2
\end{aligned}	
\end{equation}
Similarly, the MFGP built in the high-dimensional latent space can be formulated as:

\begin{equation}
\begin{small}
\begin{aligned}			
{{\bf{k}}_M}({\bf{z}},Z^*)&=
\left[ {{\rho _{11}}{\bf{k}}({\bf{z}},{Z}),{\rho _{12}}{\bf{k}}({\bf{z}},{Z'})} \right]\\
{K_M}({Z^*}) &= \left[ {\begin{array}{*{20}{c}}
{{\rho _{11}}K(Z)}&{{\rho _{12}}K(Z,Z')}\\
{{\rho _{21}}K(Z',Z)}&{{\rho _{11}}(Z')}
\end{array}} \right] + D
\end{aligned}
\end{small}
\end{equation}

\begin{equation}
\begin{aligned}	
\hat y\left( {\bf{z}} \right) &= {{\bf{k}}_M}\left( {{\bf{z}},{Z^*}} \right)K_M^{ - 1}\left( {{Z^*}} \right)\left[ \begin{array}{l}
{\bf{y}}\left( Z \right)\\
{\bf{y}}\left( C \right)
\end{array} \right]\\
{\sigma ^2}\left( {\bf{z}} \right) &= {{\bf{k}}_M}\left( {\bf{z}} \right) - {{\bf{k}}_M}\left( {{\bf{z}},{Z^*}} \right)K_M^{ - 1}{{\bf{k}}_M}\left( {{Z^*},{\bf{z}}} \right) + \sigma _{n,2}^2
\end{aligned}	
\end{equation}
Correspondingly, the EI acquisition function for the sample search in the low- and high-dimensional latent space can be formulated as:
\begin{equation}
\begin{small}
\begin{aligned}
EI_{{\bf{c}}}({\bf{c}}) &=  {(f_{\min }^{} - \hat y({\bf{c}}))\Phi \left( u_{{\bf{c}}} \right)} +  {\sigma ({\bf{c}})\phi \left( u_{{\bf{c}}} \right)}\\
u_{{\bf{c}}} &= {{({f_{\min }} - \hat y_{{\bf{c}}}({\bf{c}}))} \mathord{\left/
		{\vphantom {{({f_{\min }} - \hat y_{{\bf{c}}}({\bf{c}}))} {\sigma ({\bf{c}})}}} \right.
		\kern-\nulldelimiterspace} {\sigma_{{\bf{c}}} ({\bf{c}})}}
\end{aligned}
\end{small}
\end{equation}

\begin{equation}
\begin{small}
\begin{aligned}
EI_{{\bf{z}}}({\bf{z}}) &=  {(f_{\min }^{} - \hat y({\bf{z}}))\Phi \left( u_{{\bf{z}}} \right)} +  {\sigma ({\bf{z}})\phi \left( u_{{\bf{z}}} \right)}\\
u_{{\bf{z}}} &= {{({f_{\min }} - \hat y_{{\bf{z}}}({\bf{z}}))} \mathord{\left/
		{\vphantom {{({f_{\min }} - \hat y_{{\bf{z}}}({\bf{z}}))} {\sigma ({\bf{z}})}}} \right.
		\kern-\nulldelimiterspace} {\sigma_{{\bf{z}}} ({\bf{z}})}}
\end{aligned}
\end{small}
\end{equation}


With the above, Fig.6 shows the flowchart of our proposed GMFoO algorithm, which is also summarized in Algorithm 3.
As shown in Fig.6, when conducting optimizations over the high- and low-dimensional latent spaces simultaneously, knowledge transfer is carried out in each iteration to benefit from each other.
Therefore, our proposed GMFoO can be expected to take a good balance between solution accuracy and convergence rate for solving black-box problems.

\begin{figure*}[t!]
	\centering
	\includegraphics[width=5.5in]{"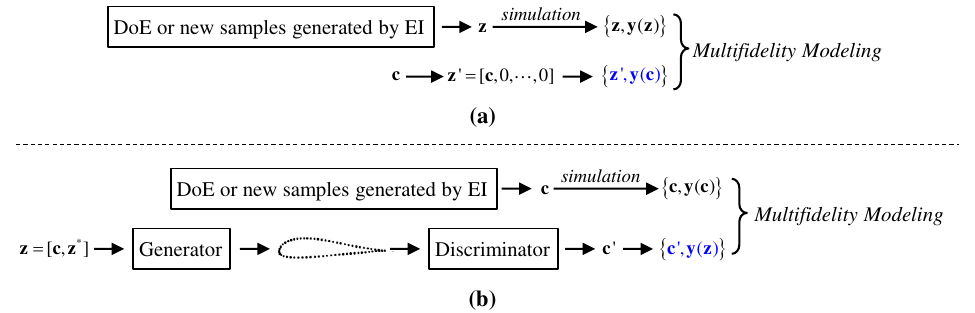"}
	\caption{Leveraging knowledge through the multi-fidelity modeling, (a) in the high-dimensional latent space, (b) in the low-dimensional latent space}
\end{figure*}


\begin{figure}
	\centering
	\includegraphics[width=3.4in]{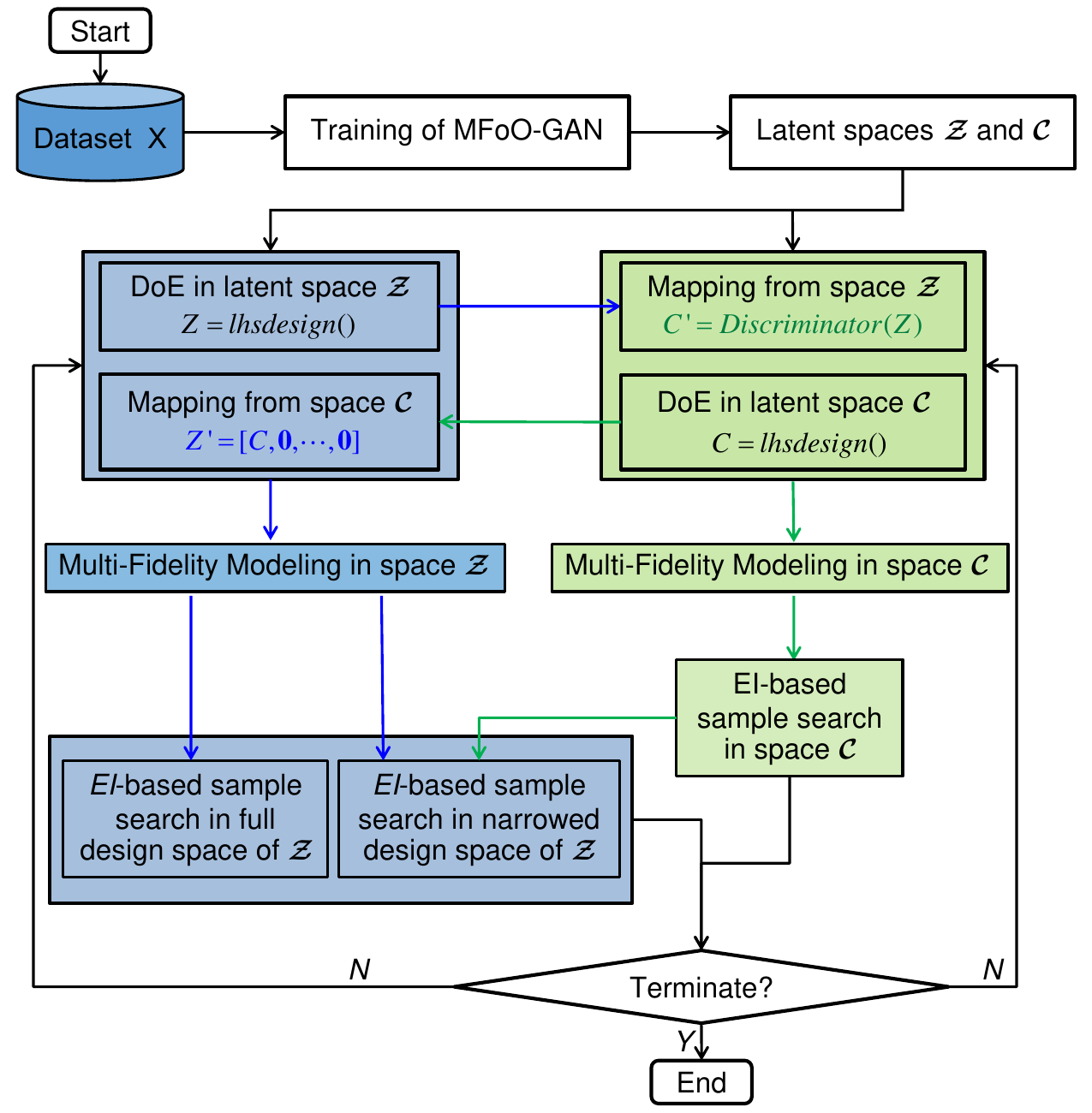}
	\caption{Flowchart of GMFoO}
\end{figure}

\begin{algorithm}[t]
	\caption{Pseudo code of GMFoO}
	\begin{algorithmic}[1]
		\REQUIRE{Target optimization problem, dataset \begin{small}$X$\end{small} with structured input space}
		\ENSURE{Optimal solution of the target problem}
		\STATE {\bf{Training of Multiple Latent Spaces:}} Train MFoO-GAN to obtain high-dimensional latent space $\mathcal{Z}$ and one or more highly correlated low-dimensional latent spaces $\mathcal{C}$ for $X$
		\STATE {\bf{DoEs in Multiple Latent Spaces:}} Use space-filling technique to generate the initial samples and evaluate related objective function to obtain the paired training sets \begin{small}$\{Z, Y(Z)\},\{C, Y(C)\}$\end{small}
		\WHILE{the termination condition is not satisfied}
		    \FOR{$i$, enumerate \begin{small}${{\mathcal{Z}}_i} \in \{ {\mathcal{Z}},{{\mathcal{C}}}\} $\end{small}}
		    \STATE {\bf{Build or Update Multi-Fidelity GP:}} Build multi-fidelity GP surrogate in latent space ${\mathcal{Z}}_i$ (see Eqs.(17) and (19))
		    \STATE {\bf{Sample Search with EI:}} Use the EI acquisition function to find the next sample to query in ${\mathcal{Z}}_i$, i.e., \begin{small}${{\bf{z}}_i^*} = \arg \mathop {\max }\limits_{{{\bf{z}}_i}} EI({\bf{z}}_i)$\end{small}
		    \STATE {\bf{Update the Training Set:}} Evaluate the label $y^*$ of the new sample \begin{small}${\bf{z}}_i^*$\end{small} and update the training set of ${\mathcal{Z}}_i$
		    \ENDFOR
		    		\STATE \textbf{Exploitation in the high-dimensional latent space:} \\
    		\quad $a$. Update the narrowed space of $\mathcal{Z}$ (see Fig.4)\\
    		\quad $b$. Sample search in the narrowed space of $\mathcal{Z}$ with EI\\
    		\quad $c$. Samples exchange between $\mathcal{Z}$ and $\mathcal{C}$ (see Fig.3)\\
		\ENDWHILE
	\end{algorithmic}
\end{algorithm}

\section{Experiments}

In this section, tests are carried out on airfoil and corbel design problems and an area maximization problem as well to show the efficacy of our proposed GMFoO.

\subsection{Baselines for comparison}

To examine the performance of GMFoO, we compare it against four kinds of algorithms as shown below:

(1) To show the advantage of multi-form optimization in GMFoO, we compare it against the conventional GMOs using single latent space. 
More specifically, the conventional GMOs are carried out with the high- and low-dimensional latent space of MFoO-GAN, respectively, and the standard BO algorithm is used as the optimizer.
The related GMO procedures are labeled as GMO-High and GMO-Low, respectively.
 
(2) To illustrate the advantage of GMFoO over traditional MFoO method, we compare it with a GMO which uses a multi-form BO algorithm (labeled as NashEGO~\cite{2018Nash}) to optimize the high-dimensional latent space of MFoO-GAN. 
The related procedure is labeled as GMO-NashEGO.
The main difference between GMO-NashEGO and GMFoO is as follows. 
The alternate searching spaces of GMO-NashEGO are generated randomly.
Differently, we promote positive correlation between the alternate searching spaces with MFoO-GAN, and thereby facilitating effective information exchange in multi-form optimization process.


(3) To show the advantage of GMFoO over non-generative method, the dimension-reduction (DR) method is also selected for the test. In particular, we follow the procedure in~\cite{2015Metric}, the singular value decomposition (SVD) approach is used to map the structured input space into a continuous latent space. Then, optimization is carried out in the related latent space with the standard BO algorithm, which is denoted by SVD-BO.

(4) To inspect the influence of optimization algorithms on the performance of GMO, the state of the art genetic algorithm CMA-ES~\cite{2003Reducing} and a recently proposed kriging-assisted evolutionary algorithm labeled as IKEA~\cite{2021A} are used for the optimization over high-dimensional latent space of MFoO-GAN.
Accordingly, the related GMO procedures are labeled as GMO-CMAES and GMO-IKEA, respectively.

\subsection{Implementing Details}

Tensorflow and PyTorch packages are used to train the MFoO-GAN, and Matlab is used to build the optimization algorithms. Then, a PERL script is programmed to build the connections between Python and Matlab codes.
More specifically, the GPML toolbox~\cite{rasmussen2010gaussian} is used to build the BO algorithms such as the standard BO, NashEGO and the optimizer of GMFoO. 
Meanwhile, the code of IKEA published by the authors in Github is used, and the code of CMA-ES is downloaded from the Mathworks\footnote{http://yarpiz.com/235/ypea108-cma-es}.

The Latin hypercube sampling~\cite{stein1987large} is used for the design of experiment (DoE) for the BO algorithms and IKEA. The number of initial training samples for the standard BO and IKEA are set as 11 times of the dimension of the low-dimensional latent space of MFoO-GAN (i.e., $11d_L$).
For the optimizer of GMFoO, the initial training samples are split into halves for the optimization in high- and low-dimensional latent space, respectively.
Note that the NashEGO and CMA-ES algorithms start the optimization 
search with a random sample. 
This starting point is set as the best solution of the DoEs of the other compared algorithms.
Meanwhile, the population size of CMA-ES and IKEA are set to be $11d_L$. 

Additionally, two parameters can be tuned in GMFoO, which are the size parameter $\Delta $ of narrowed high-dimensional latent space and the dimension of low-dimensional latent space $d_L$. 
By default, we set $\Delta = 0.15 $, and $d_L$ is varied in the range of [3, 5] in the following tests. 
After that, sensitivity analysis are carried out to discuss the influence of $\Delta $ and $d_L$ on GMFoO performance in subsection $F$.

\subsection{Tests on Airfoil Design Problems}

The airfoil shape optimization has been extensively used as a benchmark to showcase the effectiveness of newly proposed algorithms~\cite{chiu2020airfoil},~\cite{2021Guo},~\cite{wang2018hierarchical}. 
Figure 7 shows a brief introduction of the airfoil design.
The forces that acting on the airfoil can be decomposed into two components, i.e., the drag $D$ and Lift $L$, which are the functions of airfoil contours.
Meanwhile, the airfoil contour of the best design performance may vary according to the working condition that defined by the Mach number ($Ma_{\infty }$), Reynolds number ($Re$) and the angle of attack ($AoA$).

For the airfoil design optimization in this work, 192 points are used to describe the airfoil contour.
Accordingly, there are 384 design variables in this structured space.
More importantly, these design variables are highly interacted in order to generate meaningful airfoil contours. 
Instead of carrying out optimization in such high-dimensional structured space directly, we use MFoO-GAN to map it into continuous latent spaces with dozens of variables.
The UIUC dataset\footnote{https://m-selig.ae.illinois.edu/ads.html} is used as the training set.
As a comparison baseline, the non-generative model SVD is also used to train the latent space with the UIUC dataset.
After that, the performance of GMFoO is examined through the optimization of airfoils that working under low-speed and subsonic conditions, respectively. 
More detailed descriptions are given as below: 

\begin{figure}
	\centering
	\includegraphics[width=3in]{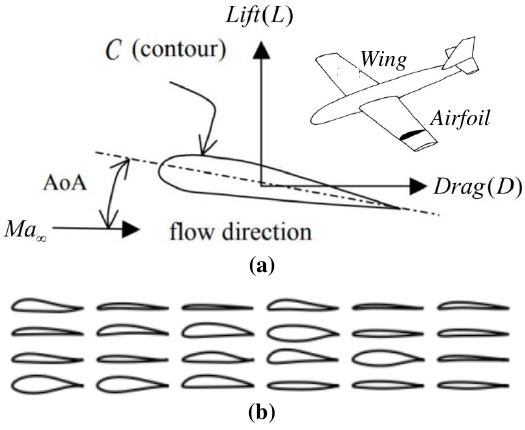}
	\caption{Introduction of (a) the airfoil contour and the forces acting on it, and (b) airfoils in UIUC database}
\end{figure}

~\\
\indent
1. Design Optimization for a Low-Speed Airfoil

The low-speed airfoils have been widely used for unmanned aerial vehicles (UAVs)~\cite{williamson2012summary}. 
Following the settings in~\cite{chen2019aerodynamic}, the design objective is to maximize the lift to drag ratio, $L/D$, with the working condition set as $Re = 1.8 \times 10^6$,
Mach number $Ma_{\infty } = 0.01$, and $AoA$ = 0 $deg$.
The XFOIL~\cite{drela1989xfoil} is used to compute $L/D$.
Meanwhile, the dimension of high- and low-dimensional latent spaces of MFoO-GAN are set to be 13 and 3, respectively. The dimension of latent space of SVD is also set to be 13. 


\begin{figure}
	\centering
	\includegraphics[width=3in]{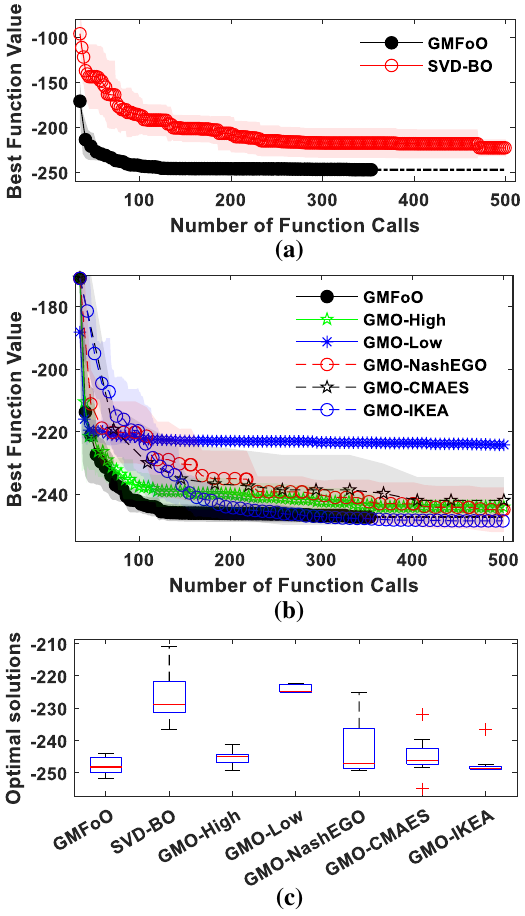}
	\caption{Testing results of low-speed airfoil, where the dimension of high- and low-dimensional latent spaces of MFoO-GAN are set to be 13 and 3, respectively}
\end{figure}



~\\
\indent
2. Design Optimization for a Subsonic Airfoil

Similar to the low-speed airfoil design, we still use $L/D$ to optimize the subsonic airfoil with XFOIL.
By refering to ~\cite{li2020efficient}, the working condition of the the target airfoil is set as follows, i.e., $Re = 1.8\times 10^6$, $Ma_{\infty } = 0.45$ and $AoA$ = 0 $deg$.
And further, to test the effectiveness of GMFoO in different settings, the dimension of high- and low-dimensional latent spaces of MFoO-GAN are set to be 23 and 4, respectively.
For SVD, the dimension of latent space is set to be 23.


\begin{figure}
	\centering
	\includegraphics[width=3in]{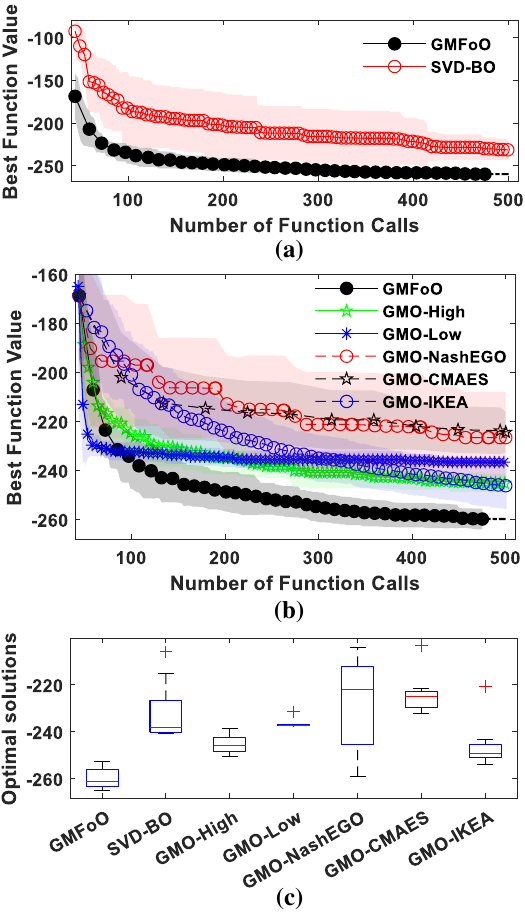}
	\caption{Testing results of subsonic airfoil, where the dimension of high- and low-dimensional latent spaces of MFoO-GAN are set to be 23 and 4, respectively}
\end{figure}

~\\
\indent
Figures 8 and 9 show the optimization results of low-speed and subsonic airfoils, where the shaded areas in Figs.8(a), 8(b), 9(a) and 9(b) show the variations of the optimization results over 10 runs.
The boxplots in Figs.8(c) and 9(c) exhibit the medians and distributions of the final optimal solutions.

As shown in Figs.8(a) and (b), both the convergence rates and final optimal solutions of GMFoO are better than those of SVD-BO, which are consistent with the observations in~\cite{chiu2020airfoil}.
The reason behind is as follows.
That is, SVD is essentially a linear dimension reduction model. 
Hence, the solution accuracy of the related latent space of SVD can be worse than those of nonlinear deep learning models such as MFoO-GAN, resulting in the poor performance of SVD-BO.

Among GMFoO, GMO-High and GMO-Low, GMO-Low converges most quickly at its optimal solution, while GMO-High always achieves better final solutions than those of GMO-Low.
In the meantime, the convergence rate of GMO-High (in 23-dimensional design space) is observed to be slower than that of GMO-Low (in 4-dimensional design space), as shown in Fig.9(b).
Differently, by carrying out optimizations in both high- and low-dimensional latent spaces simultaneously to complement each other, our proposed GMFoO takes a good balance in between solution accuracy and convergence rate, which always achieves the best solutions with even faster convergence rates.

Though carrying out multi-form optimization with alternate subspaces, the performance of GMO-NashEGO looks poorer than those optimizing over single latent space, such as GMO-High and GMO-Low.
The reason behind can be explained as follows.
Firstly, unlike GMFoO which promotes positive correlations between alternate searching spaces to facilitate effective knowledge transfer, the subspaces are generated randomly in GMO-NashEGO.
And accordingly, the knowledge transfer between alternate formulations in GMO-NashEGO may be not that efficient.
Hence, the performance of GMO-NashEGO are poor as show in our testing cases.

Among GMFoO, GMO-IKEA, GMO-CMAES and GMO-High,
the convergence rates of EA-based GMOs, i.e., GMO-IKEA and GMO-CMAES, are slower than those of BO-based GMOs, i.e., GMO-High and GMFoO.
This can be easy to be understood, as the BO algorithms are usually more efficient than the EA algorithms when solving problems in moderate dimension.
In the meantime, owing to the excellent global search performance of EA algorithms, the final solutions of IKEA looks slightly better than those of GMO-High and GMFoO, when optimizing over the 13-dimensional latent space for the low-speed airfoil (see Fig.8(b)).
However, when increasing the latent dimension from 13 to 23, as shown in Fig.9(b), GMO-High and GMFoO achieve better solutions than those of GMO-IKEA within sample budget.

\subsection{Test on Decorative Corbel Design Problems}

Corbel is a common category of decorative architectural geometry, as shown in Fig.10(a).
When designing corbels, we need to meet both aesthetics and mechanical requirements, i.e., the decorative corbel needs to look beautiful while maintaining good mechanical performance~\cite{2021Generative}.
Without loss of generality, we present a simplified mechanical model for the mainbody of the corbel as shown in Fig.10(b).
There are two forces, i.e., the gravity $mg$ and a force $F$ acting on the mainbody. 
The centroid and the center of gravity of the mainbody are denoted by $C$ and $G$, respectively.

The corbel weight and the locations of $C$ and $G$ are functions of the corbel curve. 
On one hand, to prevent the mainbody from falling off the base, the weight and the moment of gravity should be optimized to be as small as possible, making the related optimal curve be very flat. 
On the other hand, the corbel with too flat curve may not look good. 
To meet both the mechanical and aesthetic requirements, we formulate the design problem of the corbel curve as follows:
\begin{equation}
\min f\left( {\bf{x}} \right) = \;{w_1}mg{d_G} + {w_2}{\left( {{d_C} - {d_C^*}} \right)^2}
\end{equation}
where, $d_G$ and $d_C$ are distances of $G$ and $C$ to the fixed wall surface, respectively, and $d_C^*$ is the idealized centroid distance that meets the aesthetics requirement. And $w_1$ and $w_2$ are the weights of the mechanical and aesthetics objectives.

Similar to the airfoil design optimization, we use 192 points to describe the corbel curve. 
And accordingly, such structured input space for the corbel curve design has 384 design variables, which are highly interacted in order to generate meaningful corbel curves.
Therefore, instead of optimizing over such high-dimensional and discrete structured space directly, we train MFoO-GAN and SVD to map it into continuous latent spaces with dozens of variables.
The dataset from~\cite{2021Generative} is used for the training of MFoO-GAN and SVD.
The dimension of high- and low-dimensional latent spaces of MFoO-GAN are set to be 23 and 5, respectively. The dimension of latent space of SVD is also set to be 23.
Meanwhile, we set $d_C^*=(0,0)$ and have ${w_1} = 1,{w_2} = 10$ in Eq.(22).

\begin{figure}
	\centering
	\includegraphics[width=3in]{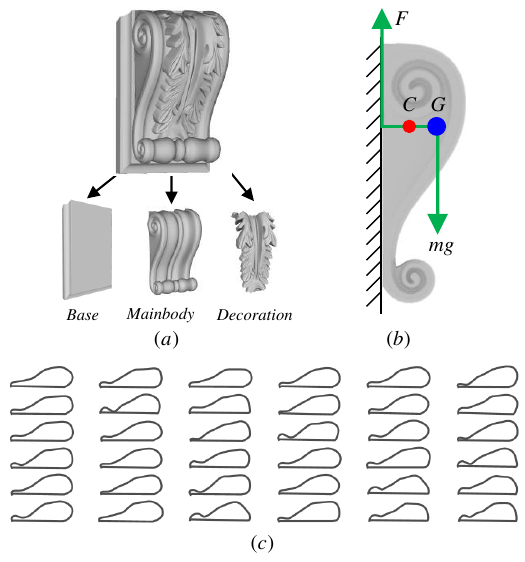}
	\caption{Introduction to the decorative corbel, (a) decorative corbel and its components, (b) forces acting on the mainbody of corbel, (c) profiles of corbel mainbody in the database}
\end{figure}

\begin{figure}
	\centering
	\includegraphics[width=3in]{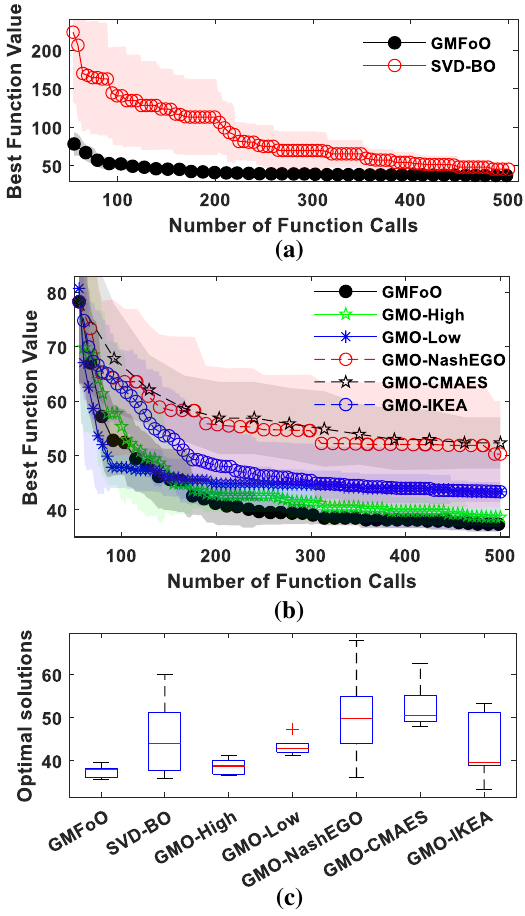}
	\caption{Testing results of corbel optimization, where the dimension of high- and low-dimensional latent spaces of MFoO-GAN are set to be 23 and 5, respectively}
\end{figure}


Figure 11 shows the optimization results of the corbel design over 10 runs.
Within sample budget, the BO-based GMOs such as GMO-High and GMFoO, achieve better solutions with faster convergence rates than those of EA-based GMOs.
When comparing between GMO-Low and GMO-High, the convergence rate of GMO-Low is faster while the final solutions of GMO-High are better.
By carrying out optimization in both high- and low-dimensional latent spaces simultaneously, our proposed GMFoO achieves a good balance in between solution accuracy and convergence rate.

\subsection{Test on Area Maximization Problem}

By following the idea in~\cite{tripp2020sample}, we further test our proposed GMFoO for finding the $28 \times 28$ binary image that has the largest total area (i.e. the largest number of pixels with value 1) in the MNIST dataset.
Note that the original design space for the above optimization problem is a 784-dimensional structured space, where the design variables can be only valued as 0 and 1.
More importantly, the design variables in the structured input are highly interacted in order to generate meaningful images.

To solve the above problem efficiently, the first step is to embed the structured input space into a continuous latent space that can generate high-quality images.
Figure 12 shows the images generated by MFoO-GAN and SVD, respectively.
Obviously, both the high- and low-dimensional latent spaces of MFoO-GAN can generate various hand-writing digits clearly. 
However, SVD cannot generate any meaningful images.
In other words, the SVD-based approach (labeled as SVD-BO in this paper) fails to solve the area maximization task of binary images.
Therefore, only the GMO approaches are compared for solving this challenging toy problem.

Figure 13 shows the optimization results of the area maximization problem.
Similar to the optimization results shown in airfoil and corbel design tasks, our proposed GMFoO achieves the best solutions with even faster convergence rate.
With the above, the effectiveness of our proposed GMFoO has been well demonstrated.


\begin{figure}
	\centering
	\includegraphics[width=3in]{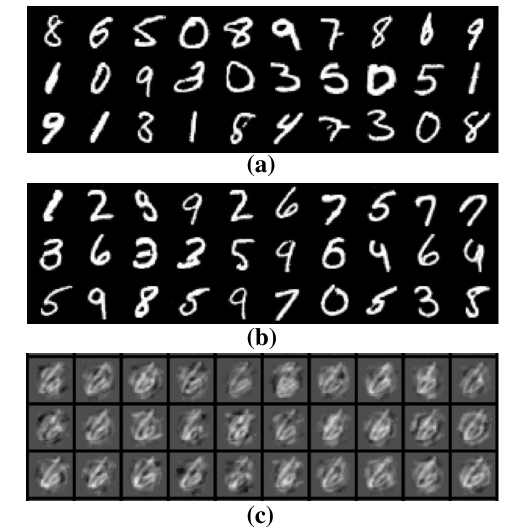}
	\caption{Generated hand-writing digits by (a) the 20-dimensional latent space and (b) the 4-dimensional latent space of MFoO-GAN, and (c) the 20-dimensional latent space of SVD}
\end{figure}
\begin{figure}
	\centering
	\includegraphics[width=3in]{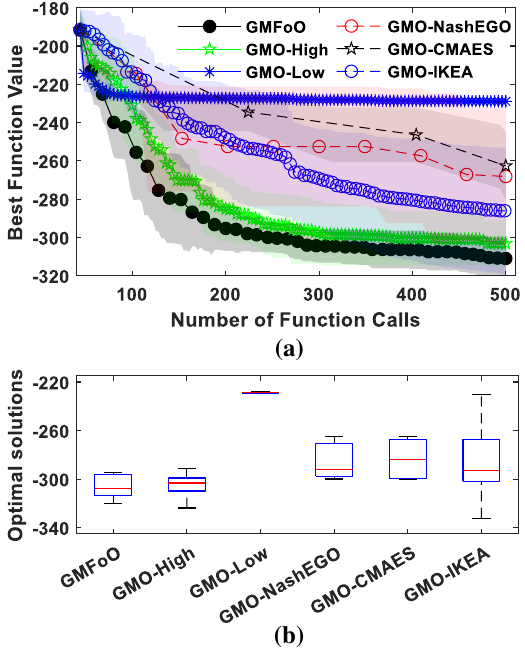}
	\caption{Testing results of area maximization problem based on the MNIST dataset, where the dimension of high- and low-dimensional latent spaces of MFoO-GAN are set to be 20 and 4, respectively}
\end{figure}

\subsection{Discussions of Algorithm Performance}

In this subsection, the reason behind the superior performance of GMFoO is discussed by analyzing the relations between the high- and low-dimensional latent spaces of MFoO-GAN.
And further, the effects of algorithm parameters on GMFoO performance are also investigated.


~\\
\indent
1. Relations between latent spaces in MFoO-GAN

Since the performance of multi-form transfer optimization is sensitive to the degree of underlying inter-task similarities~\cite{gupta2017insights}, the functional correlation between the high- and low-dimensional latent spaces of MFoO-GAN and the distributions of optimal solutions of latent spaces are analyzed.

Figure 14 shows the functional correlation between the high- and low-dimensional latent spaces. 
Specifically, recalling Section III.B, the sample set $\{ {\bf{c}}',y({\bf{z}})\}$, that transformed from the high-dimensional sample set $\{ {\bf{z}},y({\bf{z}})\}$, are used as low-fidelity samples to build the multi-fidelity GP.
Therefore, to analyze the correlation between the high- and low-dimensional latent space, the real function value of ${\bf{c}}'$, i.e., $y({\bf{c}}')$, is calculated (see the dashed path in the upper of Fig.14). 
In particular, 1000 sample pairs as $\{ y({\bf{z}}),y({\bf{c}}')\} $ are calculated to plot the scatterplot. 
Obviously, the sample pairs are closely distributed along the dashed line, and the Pearson's coefficient of sample pairs is 0.8.
It indicates that the high- and low-dimensional latent space are well correlated, and therefore accurate multi-fidelity GP surrogate can be built~\cite{guo2018analysis} to facilitate positive knowledge transfer in GMFoO.
\begin{figure}
	\centering
	\includegraphics[width=3in]{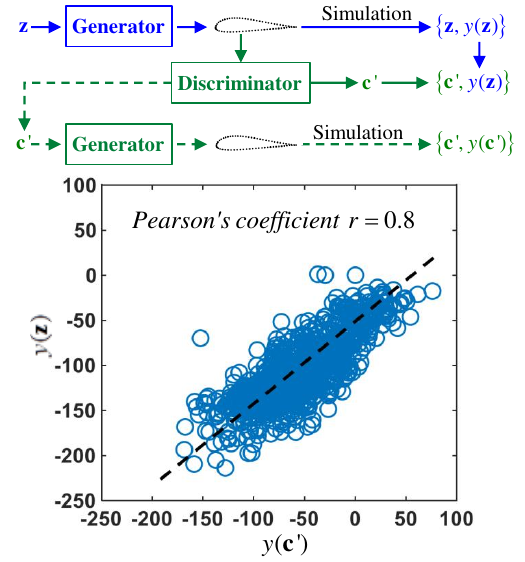}
	\caption{Functional correlation between high- and low-dimensional latent spaces of MFoO-GAN, by taking low-speed airfoil optimization as an example }
\end{figure}

\begin{figure*}
	\centering
	\includegraphics[width=6.5in]{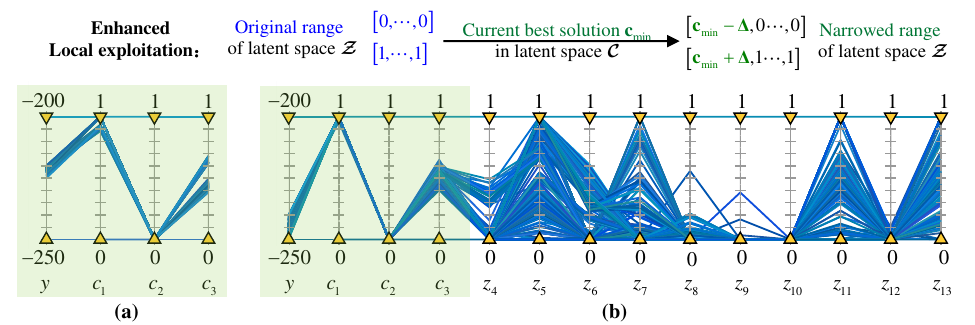}
	\caption{Distributions of the optimal solutions in the high- and low-dimensional latent spaces of MFoO-GAN, by taking low-speed airfoil optimization as an example, (a) Optimal solutions in the low-dimensional latent space, (b) Optimal solutions in the high-dimensional latent space}
\end{figure*}
Figure 15 shows the distributions of optimal solutions over 10 runs, by using the parallel coordinates.
The intersections of a polyline and vertical axes in the parallel coordinates exhibit the component parameter values of a optimal solution.
As shown in the shaded area of Figs.15(a) and (b), the distribution ranges of $c_1$, $c_2$ and $c_3$ of the optimal solutions of the high- and low-dimensional latent spaces are almost overlapped.
In the meantime, the function values of optimal solutions of the low-dimensional latent space are approaching towards those of the high-dimensional latent space. 
It confirms that the optimization in the low-dimensional latent space of MFoO-GAN does help to prioritize the search in the directions of major variability of the target object.
Then, by leveraging the optimal solution of low-dimensional latent space through the enhanced local exploitation strategy (see the upper in Fig.15 and Section III.B), the optimizer of the high-dimensional latent space can arrive at the neighborhood of the real optimum more quickly.
With the above, it is not hard to comprehend why GMFoO can always achieve better solutions than the compared algorithms as shown in Subsection D.

~\\
\indent
2. Effects of the size parameter \begin{small}$\Delta $\end{small}

Note that a multi-fidelity GP based optimization and an local exploitation strategy are proposed in GMFoO, their effectiveness are analyzed by varying the size parameter \begin{small}$\Delta $\end{small}.
Specifically, \begin{small}$\Delta $\end{small} controls the area of local exploitation (see Fig.4). 
When \begin{small}$\Delta =0 $\end{small}, only the multi-fidelity GP based optimization is conducted in GMFoO, which is denoted by GMFoO\_0. 
Further, to show the combined effects and the sensitivity of \begin{small}$\Delta $\end{small} on GMFoO performance, \begin{small}$\Delta $\end{small} is also set to be 0.1, 0.15 and 0.2, and the related GMFoO algorithms are labeled as GMFoO\_10, GMFoO\_15 and GMFoO\_20, respectively.

Figure 16 shows the testing results of low-speed and subsonic airfoil optimizations, respectively. Compared to GMO-High, GMFoO\_0 consistently achieves better results, and therefore the effectiveness of multi-fidelity GP based optimization is demonstrated.
In the meantime, the medians of GMFoO\_10, GMFoO\_15 and GMFoO\_20 are better than those of GMFoO\_0, indicating that the proposed local exploitation strategy can also help to improve the GMFoO performance.

Among GMFoO\_10, GMFoO\_15 and GMFoO\_20, the performance of GMFoO\_15 is most robust, which achieves the second best and the best results for the low-speed and subsonic airfoil optimization, respectively.  
The reason behind can be explained as follows. 
The neighborhood of optimal solutions of the high- and low-dimensional latent space cannot be exactly overlapped. Hence, too small \begin{small}$\Delta $\end{small} may lead the exploitation area get out of the vicinity of the real optimal of the high-dimensional latent space. 
On the other hand, too large \begin{small}$\Delta $\end{small} may result in worse sample efficiency and thus worse results within budget.
In other words, there should be a balance when selecting \begin{small}$\Delta $\end{small} to achieve the global optimal solution efficiently. Hence, GMFoO with $\Delta  = 0.15$ (i.e., GMFoO\_15) achieves the best solutions in our tests.

\begin{figure}
	\centering
	\includegraphics[width=3.2
	in]{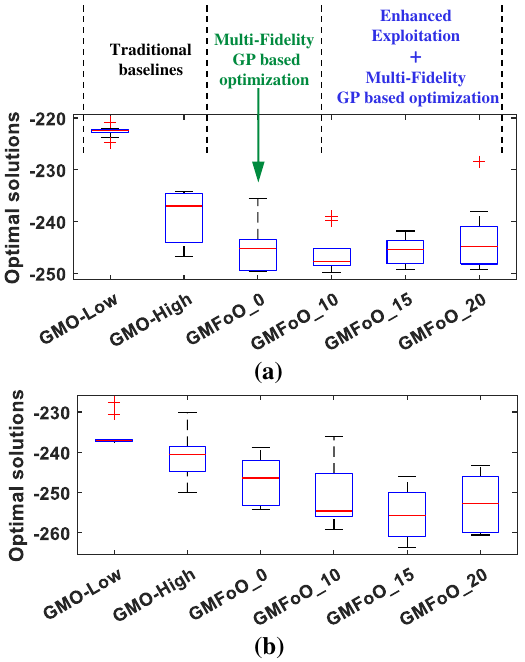}
	\caption{Experimental results with different $\Delta$ in GMFoO, (a) low-speed airfoil optimization, (b) subsonic airfoil optimization}
\end{figure}


~\\
\indent
3. Effects of the dimension of low-dimensional space
\begin{figure}
	\centering
	\includegraphics[width=3.4in]{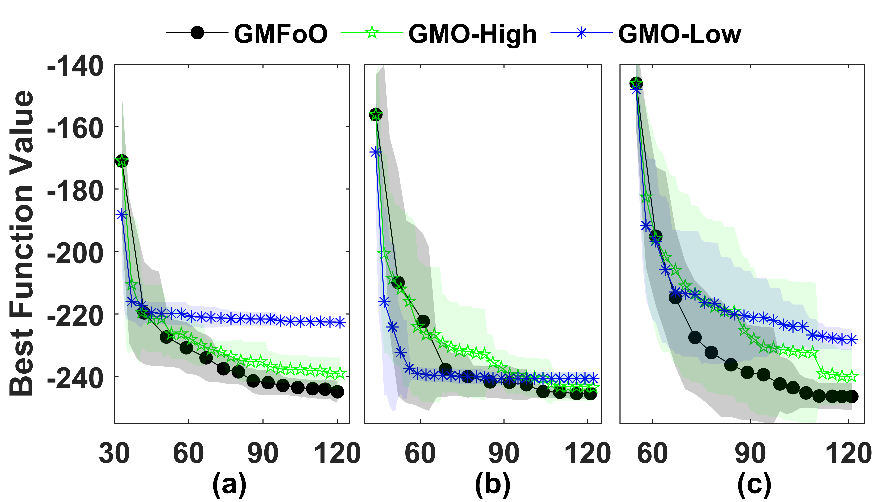}
	\caption{Effects of $d_L$ on GMFoO performance by testing on low-speed airfoil design, (a) $d_L=3$, (b) $d_L=4$, (c) $d_L=5$}
\end{figure}

\begin{figure}
	\centering
	\includegraphics[width=3.4in]{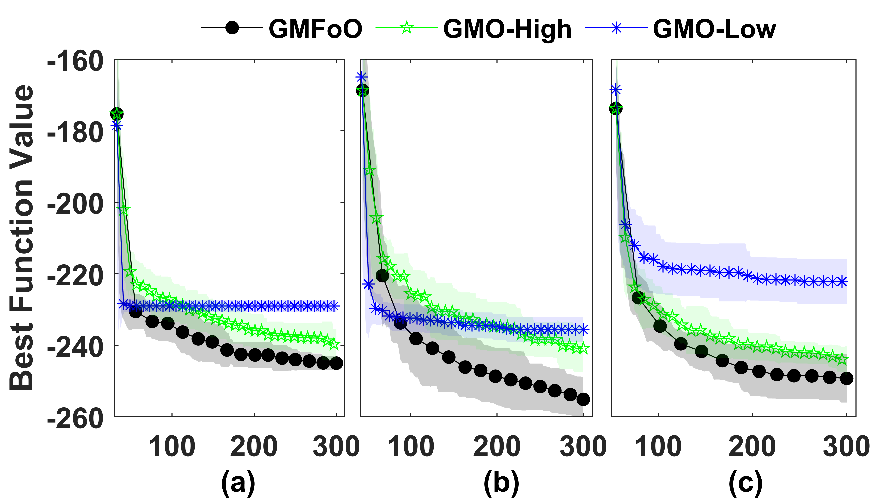}
	\caption{Effects of $d_L$ on GMFoO performance by testing on subsonic airfoil design, (a) $d_L=3$, (b) $d_L=4$, (c) $d_L=5$}
\end{figure}

To inspect the effects of the dimension of low-dimensional latent space $d_L$ on GMFoO performance, we change $d_L$ from 3 to 5 in the following tests.
According to the convergence histories of GMFoO for low-speed and subsonic airfoil optimizations shown in Figs.8 and 9, the sample budget for the low-speed and subsonic airfoil optimization is set as 120 and 300, respectively.

As the MFoO-GANs with different $d_L$ are trained separately, the landscape of related latent spaces can be different.
Hence, the convergence histories of GMO-High look different for different cases.
However, GMFoO always achieves better solutions than those of GMO-High and GMO-Low.

When comparing GMO-Low procedures with different $d_L$, the convergence rate of GMO-Low with $d_L = 3$ is the fastest for both low-speed and subsonic airfoil optimizations. 
In the meantime, the overall performance of GMFoO with $d_L=4$ is the best, which achieves the second best and the best optimal solutions for the low-speed and subsonic airfoil optimizations, respectively.
The reasons behind can be explained as follows.
With the increase of $d_L$, the solution accuracy of low-dimensional latent space becomes better. 
However, due to the ``curse of dimensionality", much more samples are required to attain the optimal solution of the corresponding low-dimensional space.
As a compromise, the GMO-Low with $d_L = 4$ performs best within sample budget.
And accordingly, the information exchange between the low-dimensional latent space with $d_L = 4$ and the high-dimensional latent space can be most efficient, making the GMFoO with $d_L = 4$ attains the overall best results.

\section{Conclusions}

We presented a novel generative model based optimization (GMO) framework, namely GMFoO, to more efficiently solve the black-box problems that involve complex structured input space.
We showed that, instead of mapping the structured input space into single latent space as most GMO studies do, 
generating multiple latent spaces and carrying out optimizations over these latent spaces simultaneously, can help to achieve a good balance in between desirable solution accuracy and convergence rate.
By instantiating GMFoO with Bayesian optimization, the effectiveness of our proposed approach has been well demonstrated through testing on airfoil and corbel design problems and an area maximization problem.

\bibliography{References}

\begin{thebibliography}{10}
\providecommand{\url}[1]{#1}
\csname url@samestyle\endcsname
\providecommand{\newblock}{\relax}
\providecommand{\bibinfo}[2]{#2}
\providecommand{\BIBentrySTDinterwordspacing}{\spaceskip=0pt\relax}
\providecommand{\BIBentryALTinterwordstretchfactor}{4}
\providecommand{\BIBentryALTinterwordspacing}{\spaceskip=\fontdimen2\font plus
\BIBentryALTinterwordstretchfactor\fontdimen3\font minus \fontdimen4\font\relax}
\providecommand{\BIBforeignlanguage}[2]{{%
\expandafter\ifx\csname l@#1\endcsname\relax
\typeout{** WARNING: IEEEtran.bst: No hyphenation pattern has been}%
\typeout{** loaded for the language `#1'. Using the pattern for}%
\typeout{** the default language instead.}%
\else
\language=\csname l@#1\endcsname
\fi
#2}}
\providecommand{\BIBdecl}{\relax}
\BIBdecl

\bibitem{li2020efficient}
J.~Li, M.~Zhang, J.~R. Martins, and C.~Shu, ``Efficient aerodynamic shape optimization with deep-learning-based geometric filtering,'' \emph{AIAA Journal}, pp. 1--17, 2020.

\bibitem{2021Hull}
A.~Serani, F.~Stern, E.~F. Campana, and M.~Diez, ``Hull-form stochastic optimization via computational-cost reduction methods,'' \emph{Engineering with Computers}, no.~2, 2021.

\bibitem{tripp2020sample}
A.~Tripp, E.~Daxberger, and J.~M. Hern{\'a}ndez-Lobato, ``Sample-efficient optimization in the latent space of deep generative models via weighted retraining,'' \emph{arXiv preprint arXiv:2006.09191}, 2020.

\bibitem{chiu2020airfoil}
K.~Chiu, M.~Fuge \emph{et~al.}, ``Airfoil design parameterization and optimization using b$\backslash$'ezier generative adversarial networks,'' \emph{arXiv preprint arXiv:2006.12496}, 2020.

\bibitem{du2020b}
X.~Du, P.~He, and J.~Martins, ``A b-spline-based generative adversarial network model for fast interactive airfoil aerodynamic optimization,'' in \emph{AIAA Scitech 2020 Forum}, 2020, p. 2128.

\bibitem{kusner2017grammar}
M.~J. Kusner, B.~Paige, and J.~M. Hern{\'a}ndez-Lobato, ``Grammar variational autoencoder,'' in \emph{International Conference on Machine Learning}.\hskip 1em plus 0.5em minus 0.4em\relax PMLR, 2017, pp. 1945--1954.

\bibitem{2017Constrained}
R.~R. Griffiths, ``Constrained bayesian optimization for automatic chemical design,'' \emph{arXiv preprint arXiv:1709.05501}, 2017.

\bibitem{chen2019aerodynamic}
W.~Chen, K.~Chiu, and M.~Fuge, ``Aerodynamic design optimization and shape exploration using generative adversarial networks,'' in \emph{AIAA Scitech 2019 Forum}, 2019, p. 2351.

\bibitem{lu2018structured}
X.~Lu, J.~Gonzalez, Z.~Dai, and N.~D. Lawrence, ``Structured variationally auto-encoded optimization,'' in \emph{International conference on machine learning}.\hskip 1em plus 0.5em minus 0.4em\relax PMLR, 2018, pp. 3267--3275.

\bibitem{nguyen2016synthesizing}
A.~Nguyen, A.~Dosovitskiy, J.~Yosinski, T.~Brox, and J.~Clune, ``Synthesizing the preferred inputs for neurons in neural networks via deep generator networks,'' \emph{Advances in neural information processing systems}, vol.~29, pp. 3387--3395, 2016.

\bibitem{luo2018neural}
R.~Luo, F.~Tian, T.~Qin, E.~Chen, and T.-Y. Liu, ``Neural architecture optimization,'' \emph{arXiv preprint arXiv:1808.07233}, 2018.

\bibitem{yang2018microstructural}
Z.~Yang, X.~Li, L.~Catherine~Brinson, A.~N. Choudhary, W.~Chen, and A.~Agrawal, ``Microstructural materials design via deep adversarial learning methodology,'' \emph{Journal of Mechanical Design}, vol. 140, no.~11, 2018.

\bibitem{gomez2018automatic}
R.~G{\'o}mez-Bombarelli, J.~N. Wei, D.~Duvenaud, J.~M. Hern{\'a}ndez-Lobato, B.~S{\'a}nchez-Lengeling, D.~Sheberla, J.~Aguilera-Iparraguirre, T.~D. Hirzel, R.~P. Adams, and A.~Aspuru-Guzik, ``Automatic chemical design using a data-driven continuous representation of molecules,'' \emph{ACS central science}, vol.~4, no.~2, pp. 268--276, 2018.

\bibitem{shan2010survey}
S.~Shan and G.~G. Wang, ``Survey of modeling and optimization strategies to solve high-dimensional design problems with computationally-expensive black-box functions,'' \emph{Structural and multidisciplinary optimization}, vol.~41, no.~2, pp. 219--241, 2010.

\bibitem{gupta2017insights}
A.~Gupta, Y.-S. Ong, and L.~Feng, ``Insights on transfer optimization: Because experience is the best teacher,'' \emph{IEEE Transactions on Emerging Topics in Computational Intelligence}, vol.~2, no.~1, pp. 51--64, 2017.

\bibitem{da2018curbing}
B.~Da, A.~Gupta, and Y.-S. Ong, ``Curbing negative influences online for seamless transfer evolutionary optimization,'' \emph{IEEE transactions on cybernetics}, vol.~49, no.~12, pp. 4365--4378, 2018.

\bibitem{zhou2020toward}
L.~Zhou, L.~Feng, K.~C. Tan, J.~Zhong, Z.~Zhu, K.~Liu, and C.~Chen, ``Toward adaptive knowledge transfer in multifactorial evolutionary computation,'' \emph{IEEE Transactions on Cybernetics}, 2020.

\bibitem{liang2020evolutionary}
Z.~Liang, H.~Dong, C.~Liu, W.~Liang, and Z.~Zhu, ``Evolutionary multitasking for multiobjective optimization with subspace alignment and adaptive differential evolution,'' \emph{IEEE Transactions on Cybernetics}, 2020.

\bibitem{li2021meta}
J.-Y. Li, Z.-H. Zhan, K.~C. Tan, and J.~Zhang, ``A meta-knowledge transfer-based differential evolution for multitask optimization,'' \emph{IEEE Transactions on Evolutionary Computation}, 2021.

\bibitem{2016Evolutionary}
B.~Da, A.~Gupta, Y.~S. Ong, and F.~Liang, ``Evolutionary multitasking across single and multi-objective formulations for improved problem solving,'' in \emph{Evolutionary Computation}, 2016.

\bibitem{zhang2021study}
L.~Zhang, Y.~Xie, J.~Chen, L.~Feng, C.~Chen, and K.~Liu, ``A study on multiform multi-objective evolutionary optimization,'' \emph{Memetic Computing}, pp. 1--12, 2021.

\bibitem{peng2018multimodal}
X.~Peng, Y.~Jin, and H.~Wang, ``Multimodal optimization enhanced cooperative coevolution for large-scale optimization,'' \emph{IEEE transactions on cybernetics}, vol.~49, no.~9, pp. 3507--3520, 2018.

\bibitem{cai2019efficient}
X.~Cai, L.~Gao, and X.~Li, ``Efficient generalized surrogate-assisted evolutionary algorithm for high-dimensional expensive problems,'' \emph{IEEE Transactions on Evolutionary Computation}, vol.~24, no.~2, pp. 365--379, 2019.

\bibitem{cheng2015social}
R.~Cheng and Y.~Jin, ``A social learning particle swarm optimization algorithm for scalable optimization,'' \emph{Information Sciences}, vol. 291, pp. 43--60, 2015.

\bibitem{wang2017committee}
H.~Wang, Y.~Jin, and J.~Doherty, ``Committee-based active learning for surrogate-assisted particle swarm optimization of expensive problems,'' \emph{IEEE transactions on cybernetics}, vol.~47, no.~9, pp. 2664--2677, 2017.

\bibitem{2018A}
P.~I. Frazier, ``A tutorial on bayesian optimization,'' \emph{arXiv preprint arXiv:1807.02811}, 2018.

\bibitem{Martinez2019Funneled}
Martinez-Cantin and Ruben, ``Funneled bayesian optimization for design, tuning and control of autonomous systems,'' \emph{IEEE Transactions on Cybernetics}, vol.~49, no.~4, pp. 1489--1500, 2019.

\bibitem{2020Good}
E.~Siivola, J.~Gonzalez, A.~Paleyes, and A.~Vehtari, ``Good practices for bayesian optimization of high dimensional structured spaces,'' \emph{arXiv preprint arXiv:2012.15471}, 2020.

\bibitem{2021Guo}
Z.~Guo, Y.~Ong, and H.~Liu, ``Calibrated and recalibrated expected improvements for bayesian optimization,'' \emph{Structural and Multidisciplinary Optimization}, 2021.

\bibitem{2020Generalizing}
A.~Tan, A.~Gupta, and Y.~S. Ong, ``Generalizing transfer bayesian optimization to source-target heterogeneity,'' \emph{IEEE Transactions on Automation Science and Engineering}, vol.~PP, no.~99, 2020.

\bibitem{kingma2013auto}
D.~P. Kingma and M.~Welling, ``Auto-encoding variational bayes,'' \emph{arXiv preprint arXiv:1312.6114}, 2013.

\bibitem{goodfellow2014generative}
I.~Goodfellow, J.~Pouget-Abadie, M.~Mirza, B.~Xu, D.~Warde-Farley, S.~Ozair, A.~Courville, and Y.~Bengio, ``Generative adversarial nets,'' in \emph{Advances in neural information processing systems}, 2014, pp. 2672--2680.

\bibitem{creswell2018generative}
A.~Creswell, T.~White, V.~Dumoulin, K.~Arulkumaran, B.~Sengupta, and A.~A. Bharath, ``Generative adversarial networks: An overview,'' \emph{IEEE Signal Processing Magazine}, vol.~35, no.~1, pp. 53--65, 2018.

\bibitem{min2017knowledge}
A.~T.~W. Min, R.~Sagarna, A.~Gupta, Y.-S. Ong, and C.~K. Goh, ``Knowledge transfer through machine learning in aircraft design,'' \emph{IEEE Computational Intelligence Magazine}, vol.~12, no.~4, pp. 48--60, 2017.

\bibitem{yogatama2014efficient}
D.~Yogatama and G.~Mann, ``Efficient transfer learning method for automatic hyperparameter tuning,'' in \emph{Artificial intelligence and statistics}, 2014, pp. 1077--1085.

\bibitem{tan2021evolutionary}
K.~C. Tan, L.~Feng, and M.~Jiang, ``Evolutionary transfer optimization-a new frontier in evolutionary computation research,'' \emph{IEEE Computational Intelligence Magazine}, vol.~16, no.~1, pp. 22--33, 2021.

\bibitem{wei2021review}
T.~Wei, S.~Wang, J.~Zhong, D.~Liu, and J.~Zhang, ``A review on evolutionary multi-task optimization: Trends and challenges,'' \emph{IEEE Transactions on Evolutionary Computation}, 2021.

\bibitem{joy2019flexible}
T.~n.~T. Joy, S.~Rana, S.~Gupta, and S.~Venkatesh, ``A flexible transfer learning framework for bayesian optimization with convergence guarantee,'' \emph{Expert Systems with Applications}, vol. 115, pp. 656--672, 2019.

\bibitem{lin2019multi}
J.~Lin, H.-L. Liu, B.~Xue, M.~Zhang, and F.~Gu, ``Multi-objective multi-tasking optimization based on incremental learning,'' \emph{IEEE Transactions on Evolutionary Computation}, 2019.

\bibitem{swersky2013multi}
K.~Swersky, J.~Snoek, and R.~P. Adams, ``Multi-task bayesian optimization,'' in \emph{Advances in neural information processing systems}, 2013, pp. 2004--2012.

\bibitem{2001Reducing}
J.~D. Knowles, R.~A. Watson, and D.~W. Corne, ``Reducing local optima in single-objective problems by multi-objectivization,'' in \emph{International Conference on Evolutionary Multi-criterion Optimization}, 2001.

\bibitem{2008Multiobjectivization}
J.~Handl, S.~C. Lovell, and J.~Knowles, ``Multiobjectivization by decomposition of scalar cost functions,'' in \emph{Parallel Problem Solving from Nature-ppsn X, International Conference Dortmund, Germany, September}, 2008.

\bibitem{2021Parallel}
Z.~Guo, Q.~Wang, L.~Song, and J.~Li, ``Parallel multi-fidelity expected improvement method for efficient global optimization,'' \emph{Structural and Multidisciplinary Optimization}, 2021.

\bibitem{guo2018analysis}
Z.~Guo, L.~Song, C.~Park, J.~Li, and R.~T. Haftka, ``Analysis of dataset selection for multi-fidelity surrogates for a turbine problem,'' \emph{Structural and Multidisciplinary Optimization}, vol.~57, no.~6, pp. 2127--2142, 2018.

\bibitem{li2020multi}
S.~Li, W.~Xing, R.~Kirby, and S.~Zhe, ``Multi-fidelity bayesian optimization via deep neural networks,'' \emph{Advances in Neural Information Processing Systems}, vol.~33, 2020.

\bibitem{2002Theoretical}
C.~Coello, ``Theoretical and numerical constraint-handling techniques used with evolutionary algorithms: a survey of the state of the art,'' \emph{Computer Methods in Applied Mechanics and Engineering}, vol. 191, no. 11–12, pp. 1245--1287, 2002.

\bibitem{2018Nash}
S.~Xu and H.~Chen, ``Nash game based efficient global optimization for large-scale design problems,'' \emph{Journal of Global Optimization}, vol.~71, 2018.

\bibitem{shahriari2015taking}
B.~Shahriari, K.~Swersky, Z.~Wang, R.~P. Adams, and N.~De~Freitas, ``Taking the human out of the loop: A review of bayesian optimization,'' \emph{Proceedings of the IEEE}, vol. 104, no.~1, pp. 148--175, 2015.

\bibitem{rasmussen2010gaussian}
C.~E. Rasmussen and H.~Nickisch, ``Gaussian processes for machine learning (gpml) toolbox,'' \emph{The Journal of Machine Learning Research}, vol.~11, pp. 3011--3015, 2010.

\bibitem{2013Parallel}
E.~Contal, D.~Buffoni, A.~Robicquet, and N.~Vayatis, ``Parallel gaussian process optimization with upper confidence bound and pure exploration,'' in \emph{Joint European Conference on Machine Learning and Knowledge Discovery in Databases}, 2013, pp. 225--240.

\bibitem{jones1998efficient}
D.~R. Jones, M.~Schonlau, and W.~J. Welch, ``Efficient global optimization of expensive black-box functions,'' \emph{Journal of Global optimization}, vol.~13, no.~4, pp. 455--492, 1998.

\bibitem{chen2016infogan}
X.~Chen, Y.~Duan, R.~Houthooft, J.~Schulman, I.~Sutskever, and P.~Abbeel, ``Infogan: Interpretable representation learning by information maximizing generative adversarial nets,'' in \emph{Advances in neural information processing systems}, 2016, pp. 2172--2180.

\bibitem{2015Metric}
D.~J. Poole, C.~B. Allen, and T.~Rendall, ``Metric-based mathematical derivation of efficient airfoil design variables,'' \emph{AIAA Journal}, vol.~53, no.~5, pp. 1349--1361, 2015.

\bibitem{2003Reducing}
N.~Hansen, S.~D. Müller, and P.~Koumoutsakos, ``Reducing the time complexity of the derandomized evolution strategy with covariance matrix adaptation (cma-es),'' \emph{Evolutionary Computation}, 2003.

\bibitem{2021A}
D.~Zhan and H.~Xing, ``A fast kriging-assisted evolutionary algorithm based on incremental learning,'' \emph{IEEE Transactions on Evolutionary Computation}, vol.~PP, no.~99, pp. 1--1, 2021.

\bibitem{stein1987large}
M.~Stein, ``Large sample properties of simulations using latin hypercube sampling,'' \emph{Technometrics}, vol.~29, no.~2, pp. 143--151, 1987.

\bibitem{wang2018hierarchical}
H.~Wang, J.~Doherty, and Y.~Jin, ``Hierarchical surrogate-assisted evolutionary multi-scenario airfoil shape optimization,'' in \emph{2018 IEEE Congress on Evolutionary Computation (CEC)}.\hskip 1em plus 0.5em minus 0.4em\relax IEEE, 2018, pp. 1--8.

\bibitem{williamson2012summary}
G.~A. Williamson, B.~D. McGranahan, B.~A. Broughton, R.~W. Deters, J.~B. Brandt, and M.~S. Selig, ``Summary of low-speed airfoil data, vol. 5,'' \emph{University of Illinois, Champaign, IL}, vol. 204, 2012.

\bibitem{drela1989xfoil}
M.~Drela, ``Xfoil: An analysis and design system for low reynolds number airfoils,'' in \emph{Low Reynolds number aerodynamics}.\hskip 1em plus 0.5em minus 0.4em\relax Springer, 1989, pp. 1--12.

\bibitem{2021Generative}
Y.~Zhang, C.~O. Chan, J.~Zheng, S.~T. Lie, and Z.~Guo, ``Generative design of decorative architectural parts,'' \emph{The Visual Computer}, no.~1, pp. 1--17, 2021.

\end{thebibliography}
\bibliographystyle{IEEEtran}
\end{document}